\documentstyle[prd,aps,graphicx]{revtex}

\draft

\def\be{\beta} 
\def\ga{\gamma}

\def\si{\sigma}

\def\La{\Lambda}

\def\bk{{\mathbf{k}}}

\def\bv{{\mathbf{v}}}
\def\bx{{\mathbf{x}}}

\def\bB{{\mathbf{B}}}

\newcommand{\ben}{\begin{equation}}
\newcommand{\een}{\end{equation}}
\newcommand{\bea}{\begin{eqnarray}}
\newcommand{\eea}{\end{eqnarray}}
\newcommand{\ba}{\begin{array}}
\newcommand{\ea}{\end{array}}
\newcommand{\bi}{\begin{itemize}}
\newcommand{\ei}{\end{itemize}}

\def\math{\mathsurround 0pt}
\def\oversim#1#2{\lower.5pt\vbox{\baselineskip0pt \lineskip-.5pt
        \ialign{$\math#1\hfil##\hfil$\crcr#2\crcr{\scriptstyle\sim}\crcr}}}

\def\pa{\partial}

\title{Magnetic fields in the early universe in the
       string approach to MHD} 

\author{Mattias Christensson\cite{mcaddress}} 
\author{Mark Hindmarsh\cite{mhaddress}}
  \address{Centre for Theoretical Physics \\
  University of Sussex \\ Brighton BN1 9QJ \\ U.K} 

\preprint{SUSX-TH-99-001, 
hep-ph/9904358}

\begin{document}

\maketitle


\begin{abstract}
There is a reformulation of magnetohydrodynamics in which the fundamental 
dynamical quantities are the positions and velocities of the lines of 
magnetic flux in the plasma, which turn out to obey equations of motion 
very much like ideal strings.   We use this approach to study the 
evolution of a primordial magnetic field generated during the radiation-dominated 
era in the early Universe.  Causality dictates that the field lines form a 
tangled random network, and the string-like equations of motion, plus the
assumption of perfect reconnection, inevitably lead to a self-similar 
solution for the magnetic field power spectrum.  We present the predicted 
form of the power spectrum, and discuss insights gained from the string 
approximation, in particular the implications for the existence or 
not of an inverse cascade.

\end{abstract} 
\pacs{Pacs numbers: 95.30.Q 98.80.C\\[3pt]
Preprint number: SUSX-TH-99-001}


\section{Introduction}
It has been observed that many galaxies and clusters of galaxies are endowed 
with a magnetic field with typical strength of order $10^{-6}$ \rm G \cite{kron}. 
The origin of 
these large scale magnetic fields is unknown. In order for magnetic fields to 
have this order of magnitude it is widely believed that an enormous 
amplification of an initial seed field must have taken place. This amplification is usually  
explained by dynamo theory which can enhance the magnetic field exponentially \cite{dyn}.
However, dynamos cannot create a magnetic field, and so in order for them to act
they require a seed. At present it is not clear whether this seed 
field has its origin from some astrophysical mechanism after recombination, 
during the epoch of galaxy formation and afterwards or whether the seed field is
of primordial origin, produced in the very early universe. In the latter 
scenario it is believed that the primordial magnetic field would have been frozen
into the highly conductive plasma as the universe expanded and cooled. Because
of the high conductivity diffusion would be small and magnetic flux conserved.
If a magnetic field was produced in the early universe and was present at the 
time of recombination it may have had a significant effect on many astrophysical
processes including the formation of galaxies and stars.

There are several ways of obtaining limits on cosmological magnetic fields. 
Limits have been obtained by Faraday rotation measurements of intergalactic
fields \cite{vallee,bla-bur-oli}. Other constraints have been obtained through the consideration
of the effects of magnetic fields on primordial nucleosynthesis 
\cite{kern-sta-vach,grasso-ruben,che oli-sch-tru} and on 
the distortion in the microwave background due to the presence of a cosmological
magnetic field \cite{bar-fer-silk,dur}.   

Even if a primordial magnetic field was to weak to be of astrophysical 
significance, it is still of principal interest to study cosmic magnetic fields
today because they can provide direct and important information about the kind
of physics that must have taken place in the early universe.
There have been quite a few mechanisms proposed for ways of producing magnetic
fields in the early universe. We will not discuss them here but refer to 
\cite{hind-ever} for a brief review.

In this work we will not be concerned with any particular model for the 
generation of primordial magnetic fields. Instead we will focus on the 
universial problem of how a primordial field, whatever its origin, will develop
as the universe evolved. In order to do this one needs to consider 
magnetohydrodynamics (MHD) in an expanding universe. 
Doing full numerical relativistic MHD simulations of the physics of the early
universe is hard and requires extensive computer memory and time.  

Greater dynamical range can be obtained by resorting to approximate methods. The 
cascade model of Ref.\ \cite{brand} is one such method, which is thought to 
reproduce well the flow of energy between wavenumbers of the full MHD equations,
at the cost of a severe truncation in the number of degrees of freedom.  It was 
found in \cite{brand} that energy was transferred from small to large scales in 
an inverse cascade, and that the correlation length of the initially random 
field increased with what looked like a small power of (conformal) time. This 
is pleasing if one wants to derive the galactic dynamo seed field from 
a primordial process, for general arguments of causality and energy conservation 
indicate that such an inverse cascade is actually necessary \cite{hind-ever}.

In this work we will be using a string model approach to relativistic MHD
to study the evolution of cosmic magnetic fields. The connection between MHD 
and string dynamics have previously been studied by Semenov \cite{semenov1} and 
Olesen \cite{olesen1}. Our approach is simular to that of Semenov \cite{semenov1}
but more general since we do not assume that there is a conserved  
particle number density. We take essentially the opposite direction of Olesen \cite{olesen1},
in that we derive string equations from MHD and not the other way around.

Once we have reduced the MHD equations to a string model, the results can be 
understood in terms of the coarsening dynamics of cosmic string networks 
\cite{vil-shell,hind-kib}.  The rate of increase of the network scale length $\xi$ 
is given by the characteristic velocity of waves on the string, in this 
case the Alfv\'en velocity, which  decreases as the magnetic field decreases 
in strength.  The string approach indicates that this decrease in strength 
is primarily due to reconnection on small scales:  small flux loops are 
continually created, transferring energy away from the network of infinitely 
long flux lines.  The transfer of energy from the large-scale field happens 
in a self-similar manner: the magnetic field power spectrum can be displayed 
as
\ben
|\bB_{\bk}|^2 \propto \tau^{-\alpha} P(k\xi(\tau)),
\een
where $\bk$ is wavenumber and $\tau$ conformal time.  
A powerful scaling argument 
due to Olesen \cite{olesen2} shows, in the limit of ideal MHD, 
that $\xi \propto \tau^{2/(n+5)}$, where 
the initial power spectrum behaves as $k^n$ at low $k$.  
Causality dictates that 
$n \ge 2$ \cite{dur} (and not $n\ge 0$ as one of us \cite{hind-ever} and 
another author \cite{son} has stated). As we violate 
the ideal condition by allowing reconnection, it is not clear that this 
is the correct power of $\tau$. 

This scaling law is our main result.  We see no sign of a true inverse cascade, 
in the sense that power is not transferred from small to large scales. If 
anything, the transfer is from large to small, and it is only because energy 
is being lost faster from small scales that we see an increase in the scale 
length $\xi$.


\section{Relativistic MHD and strings}
In this work we will concentrate on the ideal limit of MHD. This means that we 
neglect any viscous effects and treat matter as a perfect fluid. This is a good 
approximation at sufficiently large scales. During the radiation 
dominated era, which we are mainly concerned with here, the universe was a very
good conductor \cite{turn-widr,aho-enq}. We therefore consider the 
$\sigma\rightarrow\infty$ limit of MHD where magnetic diffusion can be ignored
and the magnetic field can be considered to be frozen into the plasma and thus
conserving magnetic flux.

The starting point for ideal relativistic MHD is the energy-momentun tensor 
\begin{equation}T^{\mu\nu}=(\rho c^{2}+p)\frac{U^{\mu}U^{\nu}}{c^{2}}-pg^{\mu\nu}+\frac{1}{4\pi}({F^{\mu}}_{\gamma}F^{\nu\gamma}-\frac{1}{4}g^{\mu\nu}F_{\gamma\delta}F^{\gamma\delta}) \label{ideal_e.m.tensor}  \end{equation}
consisting of the ideal fluid part and electromagnetic part of the energy-momentum tensor. 
Here $p$ is the fluid pressure, $\rho c^{2}$ is the energy density of the fluid, $U^{\mu}$ 
is the four-velocity of the fluid satisfying the normalisation condition 
$U^{\mu}U_{\mu}=c^{2}$ and $F^{\mu\nu}$ is the electromagnetic field tensor.

The evolution equations for the system are given by
\begin{eqnarray}
{T^{\mu\nu}}_{;\nu}&=&0  \label{cov} \\
{F^{\mu\nu}}_{;\nu}&=&\frac{4\pi}{c}J^{\mu} \label{max1}  \\
^*\!{{F}^{\mu\nu}}_{;\nu}&=&0  \label{max2}  \\
\sigma F^{\mu\nu}U_{\nu}&=&J^{\mu}-J^{\nu}U_{\nu}U^{\mu} \label{ohm}  
\end{eqnarray}
where equation (\ref{cov}) expresses covariant energy-momentum conservation, 
equations (\ref{max1}) and (\ref{max2}) are Maxwell's equations with $J^{\mu}$ being the 
four-current density. In equation (\ref{max2}) $^*\!{{F}^{\mu\nu}}$ is the dual
field tensor defined through the relation
\begin{equation}^*\!{{F}_{\mu\nu}}=\frac{1}{2}\epsilon_{\mu\nu\gamma\rho}F^{\mu\nu}  \label{dual}  \end{equation} 
where $\epsilon_{\mu\nu\gamma\rho}$ is the Levi-Cevita symbol.
Equation (\ref{ohm}) is the relativistic version of Ohm's law where
$\sigma$ is the conductivity of the fluid, measured in the fluid rest frame.

We now repeat the derivation in Subramanian and Barrow \cite{sub-bar} to show that the 
evolution equations are conformally invariant and the 
evolution can therefore be transformed from the expanding universe to a flat
(Minkowski) spacetime. In so doing we obtain an equivalent set of equations 
which are easier to handle.
Two metrics $g_{\mu\nu}$ and $\tilde{g}_{\mu\nu}$ are said to be conformally 
related to each other if $\tilde{g}_{\mu\nu}=\Omega^{2}g_{\mu\nu}$ where 
$\Omega$ is a non-zero differentiable function.

The flat Robertson-Walker line element has the form
\begin{equation}ds^{2}=dt^{2}-a^{2}(t)d\mathbf{x}^{2} \label{rw} \end{equation}
where $t$ is the comoving proper time and $a(t)$ is the scale factor.

This metric describes a isotropic and homogeneous universe with zero curva\-ture.
The appearence of a hypothetical primordial magnetic field of some strength need
not violate the assumption of isotropy and homogeniety because although the
presence of a magnetic field will locally generate bulk motions in the fluid, if
we look at sufficiently large scales isotropy and homogeniety will be regained.
At large scales the magnetic field, whatever its origin, can be considered as
essentially random since the correlation length of the field is bounded from
above by not exceeding the causal horizon. This justifies the use of the    
Robertson-Walker metric.

We introduce conformal time $\tau$ defined by
$d\tau=a^{-1}dt$ so that equation (\ref{rw}) becomes 
\begin{equation}ds^{2}=a^{2}(\tau)(d\tau^{2}-d\mathbf{x}^{2}) \end{equation}
and hence $\eta_{\mu\nu}=\tilde{g}_{\mu\nu}=\Omega^{2}g_{\mu\nu}$ 
with $\Omega=a^{-1}(\tau)$.

We note that under conformal transformations the ideal energy-momentum tensor 
$T^{\mu\nu}$ transformation as $T^{\mu\nu}=a^{-6}\tilde{T}^{\mu\nu}$. That this
is so can be seen directly from the definition of the energy-momentum tensor 
\begin{equation}T^{\mu\nu}=\frac{2}{\sqrt{-g}}\frac{\partial}{\partial g_{\mu\nu}}(\sqrt{-g}L_{matter})  \end{equation}

The new scaled fields (denoted by tilde) obey ordinary energy-momentum conservation. 
To see this we note that the ideal energy-momentum tensor is traceless, 
$T\equiv {T^{\mu}}_{\mu}=0$, provided the perfect fluid has the equation of state
$p=\frac{1}{3}\rho c^{2}$. For most of the period before decoupling, the early
universe was radiation dominated and one can use the above equation of state. 
We have
\begin{equation}{T^{\mu\nu}}_{;\nu}={T^{\mu\nu}}_{,\nu}+\Gamma^{\mu}_{\nu\rho}T^{\nu\rho}+\Gamma^{\nu}_{\sigma\nu}T^{\sigma\mu} \end{equation}
Using
\begin{equation}\Gamma^{\mu}_{\sigma\rho}=\frac{\dot{a}}{a}(\delta^{\mu}_{\sigma}\delta^{0}_{\rho}+\delta^{\mu}_{\rho}\delta^{0}_{\sigma}-g_{\sigma\rho}\delta^{\mu}_{0})\quad \textrm{and} \quad \Gamma^{\mu}_{\sigma\rho}=4\frac{\dot{a}}{a}(\delta^{\mu}_{\sigma})   \end{equation}
we get
\begin{equation}{T^{\mu\nu}}_{;\nu}={T^{\mu\nu}}_{,\nu}+2\frac{\dot{a}}{a}T^{\mu 0}-\frac{\dot{a}}{a}\delta^{\mu}_{0}T+4\frac{\dot{a}}{a}T^{\mu 0}  \end{equation}
But since $T=0$ then 
\begin{equation}(a^{-6}\tilde{T}^{\mu 0})_{,\nu}+6\frac{\dot{a}}{a}(a^{-6}\tilde{T}^{\mu 0})=({\tilde {T}}^{\mu\nu}{}_{,\nu})a^{-6}=0  \end{equation}
and hence 
\begin{equation}{\tilde {T}}^{\mu\nu}{}_{,\nu}=0    \end{equation}

This means that under conformal transformations our original variables will 
transform to a set of new scaled variable satisfying the following relations:
$\rho=a^{-4}\tilde{\rho}$, $p=a^{-4}\tilde{p}$, 
$U^{\mu}=a^{-1}\tilde{U}^{\mu}$, $J^{\mu}=a^{-4}\tilde{J}^{\mu}$,
$F^{\mu\nu}=a^{-4}\tilde{F}^{\mu\nu}$.

Now consider Maxwell's equation (\ref{max1}). 
Since $F^{\mu\nu}$ is anti symmetric the left hand side simplifies to
\begin{equation}{F^{\mu\nu}}_{;\rho}={F^{\mu\nu}}_{,\rho}+\Gamma^{\nu}_{\sigma\nu}F^{\mu\sigma}=a^{-4}{\tilde {F}}^{\mu\nu}{}_{,\rho} \end{equation}
So the equation for the scaled fields becomes
\begin{equation}{\tilde {F}}^{\mu\nu}{}_{,\nu}=\frac{4\pi}{c}\tilde{J}^{\mu} \end{equation}

For the four-velocity we have 
\begin{equation}U_{\nu}=g_{\nu\rho}U^{\rho}=a^{2}\tilde{g}_{\nu\rho}(a^{-1}\tilde{U}^{\rho})=a\tilde{U}_{\nu}  \end{equation}
Hence
\begin{equation}\sigma F^{\mu\nu}U_{\nu}=\sigma a^{-3}\tilde{F}^{\mu\nu}\tilde{U}_{\nu}=a^{-4}(\tilde{J}^{\mu}+\tilde{J}^{\nu}\tilde{U}_{\nu}\tilde{U}^{\mu})  \end{equation}
So Ohm's law remains invariant under conformal transformations if we define the 
scaled conductivity through $\sigma=\tilde{\sigma}a^{-1}$

So we arrive at the fundamental equations of relativistic MHD, 
\begin{eqnarray}
{\tilde {T}}^{\mu\nu}{}_{,\nu}&=&0   \\
{\tilde {F}}^{\mu\nu}{}_{,\nu}&=&\frac{4\pi}{c}\tilde{J}^{\mu}   \\
^*\!{\tilde {F}}^{\mu\nu}{}_{,\nu}&=&0   \\
\tilde{\sigma}\tilde{F}^{\mu\nu}U_{\nu}&=&\tilde{J}^{\mu}-\tilde{J}^{\nu}\tilde{U}_{\nu}\tilde{U}^{\mu}/c^2   
\end{eqnarray}
From here on we drop the tilde, on the understanding that we mean scaled fields.

We will now introduce a new set of coordinates which will enable us to write the
MHD equations as non-linear string equations. 
We define a magnetic four-vector $b^{\mu}$ through the relation
\begin{equation}b^{\mu}=\, ^*\!{F}_{\mu\nu}U^{\nu}/c  \end{equation}
We also define new coordinates $x'=(\eta,\sigma,\psi,\zeta)$ such that 
$\eta$ are coordinate lines of fluid elements and $\sigma$ are coordinate lines
of magnetic flux.
Hence $U'^{\mu}=(cq,0,0,0)$ and $b'^{\mu}=(0,\beta,0,0)$ satisfying 
$U'^{2}=U^{2}=c^{2}$ and $b'^{2}=b^{2}=-B^{2}$ respectively.
Thus we have the metric tensor in the new coordinates
\[g'_{\mu\nu}=diag(1/q^{2},-B^{2}/\beta^{2},-h_{AB}) \]
where $A,B=2,3$.
The new coordinate vectors are 
\begin{equation}\frac{\partial x^{\mu}}{\partial\eta}=\frac{U^{\mu}}{q}  \end{equation} 
\begin{equation}\frac{\partial x^{\mu}}{\partial\sigma}=\frac{b^{\mu}}{\beta}  \end{equation}

Since the introduction of these coordinates relies on the frozen in property
of the plasma we will refer to them as frozen-in coordinates.
In the frozen-in coordinate system we can trace the trajectories of fluid 
elements by simply varying the value of our time coordinate $\eta$ and keeping
the values of the other coordinated fixed. Similarly, we can trace the magnetic 
field lines in the frozen-in system by varying the value of $\sigma$ and keeping
the other three coordinates at fixed values.

The analysis of the MHD equations is usually performed in terms of the magnetic
field and velocity distributions. However, in the description of MHD phenomena,
the concept of a magnetic flux tube is often introduced. The reason is that it
is sometimes convenient, and we gain a better physical insight, if we base the
description on this concept rather on the magnetic field and velocity 
distributions. A magnetic flux tube is defined as the volume $V$ enclosed by a 
closed surface $\Sigma$ which is everywhere parallel to the ambient magnetic
field vector, and two cross-sectional surface areas $S_{A}$ and $S_{B}$ at 
either end. The flux tube therefore consists of a bundle of magnetic field 
lines which enter and exit the volume through the end of surfaces $S_{A}$ and
$S_{B}$.

The Cartesian coordinate system, traditionally employed in MHD does not really
lend itself to an analysis of the magnetic flux tube bahaviour. The frozen-in 
coordinate system on the other hand, does. The frozen-in coordinates provide a 
coordinate system co-moving with a flux tube, and is therefore the more 
natural choice for the mathematical analysis of flux tube behaviour.

We now consider the equation for magnetic evolution
\begin{equation}^*\!{F^{\mu\nu}}_{;\nu}=0  \end{equation}
Since $^*\!{F}^{\mu\nu}$ is antisymmetric the divergence is given by
\begin{equation}\frac{1}{\sqrt{-g}}\partial_{\nu}(\sqrt{-g}\,^*\!{F^{\mu\nu}})=0  \end{equation}
Using the fact that we can express the dual field tensor as
\begin{equation}^*\!{F}^{\mu\nu}=(b^{\mu}U^{\nu}-U^{\mu}b^{\nu})/c \label{F_b_U}\end{equation}
we have
\begin{equation}\frac{1}{\sqrt{-g}}\partial_{\nu}\left(\sqrt{-g}q\beta\Big(\frac{b^{\mu}}{\beta}\frac{U^{\nu}}{q}-\frac{U^{\mu}}{q}\frac{b^{\nu}}{\beta}\Big)\right)=0 \end{equation}
which gives
\begin{eqnarray}
\lefteqn{q\beta\left(\Big(\frac{U^{\nu}}{q}\Big)\partial_{\nu}\Big(\frac{b^{\mu}}{\beta}\Big)-\Big(\frac{b^{\nu}}{\beta}\Big)\partial_{\nu}\Big(\frac{U^{\mu}}{q}\Big)\right)+{} }\nonumber\\
& &{}\frac{1}{\sqrt{-g}}\left(\frac{b^{\mu}}{\beta}\partial_{\nu}(\sqrt{-g}\beta U^{\nu})-\frac{U^{\mu}}{q}\partial_{\nu}(\sqrt{-g}q b^{\nu})\right)=0  
\end{eqnarray}

But first square bracket is the Lie derivative and so vanishes, giving
\begin{equation}\frac{b^{\mu}}{\beta}\partial_{\nu}(\sqrt{-g}\beta U^{\nu})-\frac{U^{\mu}}{q}\partial_{\nu}(\sqrt{-g}q b^{\nu})=0  \end{equation}
Thus we have
\begin{equation}\partial_{\nu}(\sqrt{-g}\beta U^{\nu})=0  \end{equation}
and
\begin{equation}\partial_{\nu}(\sqrt{-g}q b^{\nu})=0  \end{equation}
Hence in our comoving frame we get
\begin{equation}\frac{\partial}{\partial\eta}(\sqrt{-g}\beta q)=\frac{\partial}{\partial\eta}(B\det(h_{AB}))=0  \end{equation}
\begin{equation}\frac{\partial}{\partial\sigma}(\sqrt{-g}\beta q)=\frac{\partial}{\partial\sigma}(B\det(h_{AB}))=0  \end{equation}
And so we see that
\[B\det(h_{AB})=F(\psi,\zeta)  \]
where $F(\psi,\zeta)$ is an arbitrary function of $\psi,\zeta$.
We therefore have the freedom to choose $\det(h_{AB})$ such that $F(\psi,\zeta)=\bar{B}$
where $\bar{B}$ is a constant and so
\[h_{AB}=\frac{\bar{B}}{B}\delta_{AB}  \]
which means that we can write
\[\sqrt{-g}=\frac{\bar{B}}{\beta q}  \]

We now study the equations of motion, starting from the energy-momentum tensor
given in equation (\ref{ideal_e.m.tensor}). Using the above expression, equation 
(\ref{F_b_U}), for the dual field tensor 
and the connection between the field tensor and its dual, equation (\ref{dual}), 
we can write the energy-momentun tensor in the following form   
\begin{equation}T^{\mu\nu}=\left(\rho c^{2}+p+\frac{B^{2}}{4\pi}\right)\frac{U^{\mu}U^{\nu}}{c^{2}}-\frac{b^{\mu}b^{\nu}}{4\pi}-g^{\mu\nu}\left(p+\frac{B^{2}}{8\pi}\right)  \end{equation}
From energy-momentum conservation we have
\begin{eqnarray}
{T^{\mu\nu}}_{,\nu}&=&\frac{(\rho c^{2}+p+B^{2}/4\pi))}{\beta}\frac{U^{\mu}}{c}\partial_{\nu}(\beta U^{\nu}) 
\nonumber\\
&+&\beta(U\cdot\partial)\left(\frac{(\rho c^{2}+p+B^{2}/4\pi)}{\beta}
\frac{U^{\mu}}{c}\right)-   \nonumber\\
& &{}\frac{1}{4 \pi q}\partial_{\nu}(q b^{\nu})-\frac{q}{4\pi}
(b\cdot\partial)\left(\frac{b^{\mu}}{q}\right)-g^{\mu\nu}P_{,\nu}=0  
\end{eqnarray}
Here $P$ is the total pressure from both fluid and electromagnetic field.
Note that the first and the third terms in this equation vanish. 
Writing this equation in our comoving reference frame we find
\begin{equation}
\beta q\left[\frac{\partial}{\partial\eta}
\left(\Big(\rho c^{2}+p+\frac{B^{2}}{4\pi} \Big)\frac{q}{c\beta}
\frac{\partial x^{\mu}}{\partial\eta}\right)-
\frac{1}{4\pi}\frac{\partial}{\partial\sigma}\left(\frac{\beta}{q}\frac{\partial x^{\mu}}
{\partial\sigma}\right)\right]-g^{\mu\nu}P_{,\nu}=0 \label{stringeq} 
\end{equation} 
Equation (\ref{stringeq}) is the equation of motion in the frozen-in coordinates.
In the frozen-in coordinate system the MHD equation of motion reduces to a set
of non-linear string equations. The behaviour of a magnetic flux tube is 
therefore formally analogous to that of a non-linear string.
The last term of the left hand side of equation (\ref{stringeq}) take account of 
inhomogeneity (i.e.\ pressure gradients) and it describes the coupling between 
neighbouring flux tubes whilst moving through a non-uniform plasma medium.

To summarize, we have shown that the behaviour of a magnetic flux tube is 
formally analogous to that of a string and one can therefore model a magnetised 
plasma as a fluid composed of strings. We have relied heavily on the 
arguments of Semenov and Semenov and Berkinov \cite{semenov1}, with one 
improvement: we have dropped their assumption that there is a conserved 
particle number density, which is neither necessary nor generally applicable 
in the early Universe.  Our derivation is also complementary to that of 
Olesen \cite{olesen1}, who starts from the relativistic string equations 
and shows that they can be interpreted as describing the motion of narrow 
flux tubes, providing the total pressure remains constant across the tube.


\section{Approximate string equations}
Exploiting the freedom to change coordinates in the $\si,\eta$ subspace,
we choose $\beta$ such that $\beta=qB$ and using this in the equation 
of motion (\ref{stringeq}) we have
\begin{equation}
\frac{1}{B}\frac{\partial}{\partial\eta}\left(\frac{c\,B}{v_{A}^{2}}
\frac{\partial x^{\mu}}{\partial\eta}\right)-\frac{1}{B}
\frac{\partial}{\partial\sigma}\left(B\frac{\partial x^{\mu}}
{\partial\sigma}\right)-\frac{4\pi}{B^{2}q^{2}}g^{\mu\nu}P_{,\nu}=0, 
\label{stringeq2} 
\end{equation} 
where we have defined the relativistic Alfv\'en velocity as 
\begin{equation}
v_{A}=\frac{c\,B/\sqrt{4\pi}}{\left(\rho c^{2}+p+B^{2}/4\pi\right)^{1/2}}. 
\label{alfven}
\end{equation}
Rearranging equation (\ref{stringeq2}) we can write it as
\begin{equation}
\frac{1}{v_{A}}\frac{\partial}{\partial\eta}\left(\frac{1}{v_{A}}\frac{\partial x^{\mu}}{\partial\eta}\right)-
\frac{\partial^{2}x^{\mu}}{\partial\sigma^{2}}=
-\frac{1}{B}\frac{\partial}{\partial\eta}\left(\frac{c\,B}{v_{A}}\right)
\frac{\partial x^{\mu}}{\partial\eta}+
\frac{1}{B}\frac{\partial B}{\partial\sigma}
\frac{\partial x^{\mu}}{\partial\sigma}+
\frac{4\pi}{B^{2}q^{2}}g^{\mu\nu}P_{,\nu} 
\label{stringeq3} 
\end{equation}
We now argue that for the particular situation we are interested in, it is 
justified to neglect the three terms on the right hand side of equation (\ref{stringeq3}). 
The third term on the RHS of equation (\ref{stringeq3}) can in general not be 
neglected because in many astrophysical situations pressure gradients are 
important. However, in the early universe the pressure from the magnetic field should be
much smaller than the fluid radiation pressure and since the early universe had a very 
high degree of homogeniety it follows that the gradients of the total pressure 
are small.

The second term on the RHS of equation (\ref{stringeq3}) does not have have a  
definite sign and so if averaged over time it will be zero. Dropping this term 
is equivalent to replacing $B$ by its root mean square value.

Again using the fact that the fluid radiation pressure in the early universe 
was much larger than the pressure from the magnetic field and remembering 
expression (\ref{alfven}) for the relativistic Alfv\'en velocity it is seen that
the first term on the right side of equation (\ref{stringeq3}) is indeed small
since the ratio $B/v_{A}$ is approximatly constant, thus justifying our decision
to neglect it.

We now rescale the time parameter $\eta$ to $\bar{\eta}$ through the relation
\begin{equation}
\frac{\partial}{\partial\bar{\eta}}=\frac{c}{v_{A}(\eta)}
\frac{\partial}{\partial\eta},
\end{equation}
where we have called attention to the fact that the Alfv\'en velocity may 
be time-dependent.   
Using the above mentioned approximations and our new time parameter $\bar{\eta}$
we are left with the following equation of motion 

\begin{equation}
\frac{1}{c^{2}}\frac{\partial^{2}x^{\mu}}{\partial\bar{\eta}^{2}}-
\frac{\partial^{2}x^{\mu}}{\partial\sigma^{2}}=0. \label{e:str_eqn}
\end{equation}
We also recall that the four-dimensional orthogonality of the coordinates, and 
the coordinate choice enforced by $\be = qB$,
supplements this equation with constraints
\ben
\dot x \cdot x' = 0, \qquad \frac{1}{c^2}\dot x^2 + x^{'2} = 0,
\label{e:str_con}
\een
where dot (prime)
denotes derivative with respect to the timelike parameter $\bar{\eta}$ $(\sigma)$.

We note that the equation (\ref{e:str_eqn}) and constraint (\ref{e:str_con}) is 
of exactly the same form as the equation for a Nambu-Goto string in Minkowski 
spacetime expressed in the conformal gauge \cite{vil-shell,hind-kib}.  With 
ideal Nambu-Goto strings in 
Minkowski spacetime one can further take $x^0 = c\bar{\eta}$, to obtain the system 
\bea
\frac{1}{c^{2}}\ddot{\mathbf{x}}-\mathbf{x}''&=&0,\label{codeeq} \\
 \dot{\mathbf{x}}\cdot\mathbf{x}' &=& 0  \label{codecon1} \\
 \frac{1}{c^{2}}\dot{\mathbf{x}}^{2}+\mathbf{x}^{'2} &=& 1  \label{codecon2}
\eea
This is not possible in general for MHD strings, for one can easily verify that 
\bea
\dot{\mathbf{x}}\cdot\mathbf{x}' &=& c\frac{\ga^2}{q^2}\frac{\bv \cdot \bB}{v_A B},\\
\frac{1}{c^{2}}\dot{\mathbf{x}}^{2}+\mathbf{x}^{'2} &=&  \frac{\ga^2}{q^2}
\left( \frac{\bv^2}{v_A^2} + \frac{(\bv \cdot \bB)^2}{c^2B^2}\right).
\eea
where $\gamma=(1-\mathbf{v}/c)^{-1/2}$ is the usual relativistic gamma factor.
However, as long as ${\bv \cdot \bB} = 0$ in the initial conditions, the 
first constraint is preserved by the evolution. Furthermore, we can use our 
remaining coordinate freedom to define $q$ by 
\ben
q^2 = \ga^2 \frac{\bv^2}{v_A^2},
\label{e:qchoice}
\een
in which case we really do reproduce the Nambu-Goto equations.  We should bear 
in mind however that we have made several approximations on the way, which are 
worth reiterating.
\begin{enumerate}
\item We have neglected pressure gradients.
\item We have made a kind of mean-field approximation in replacing the magnetic field 
by its root mean square value.
\item We have neglected the effect of the time-dependence of $B/v_A$.
\end{enumerate}
We have argued that all this approximations are reasonable in the context 
of the early Universe, and find that the Nambu-Goto equations can 
be used to approximate a class of MHD velocity and magnetic field configurations 
which satisfy $\bv \cdot \bB = 0$.  They may also be a 
reasonable approximation to other configurations, 
provided we understand that the constraints are satisfied in an average sense.


\section{Algorithm, simulations and results}


The equation of motion (\ref{codeeq}) can be evolved using the Smith-Vilenkin 
algorithm \cite{smith-vil}.
The Smith-Vilenkin algorithm provides an exact discrete evolution for a string network
defined on a face-centered cubic lattice and the evolution equations are
\begin{equation}\mathbf{x}(\eta+\delta,\sigma)=\frac{1}{2}[\mathbf{x}(\eta,\sigma+\delta)+\mathbf{x}(\eta,\sigma-\delta)]+\mathbf{v}(\eta,\sigma)\delta  \end{equation}
\begin{equation}\mathbf{v}(\eta+\delta,\sigma)=\frac{1}{2}[\mathbf{v}(\eta,\sigma+\delta)+\mathbf{v}(\eta,\sigma-\delta))]+\frac{1}{4\delta}[\mathbf{x}(\eta,\sigma+2\delta)-2\mathbf{x}(\eta,\sigma)+\mathbf{x}(\eta,\sigma-2\delta)]  \end{equation}
By using a leap-frog method, updating the positions and velocities at alternate timesteps,
the above equations allow us to calculate the exact future evolution of 
$\mathbf{x}$ and $\mathbf{v}$ from some appropriately choosen initial conditions.


Initial string configurations are generated by a method due to Vachaspati and
Vilenkin \cite{vach-vil}. They considered string formation in a global $U(1)$- 
model. The $U(1)$-manifold is discretized by allowing the phase to take only 
three possible values. These values are then placed randomly on the sites of the 
cubic lattice. As we go around the face of a cube in real space, the phase 
descibes a certain trajectory in the group space. A string passes through the 
face of a cube if that trajectory has a non-zero winding number. With this method
the string segments join up to form either closed loops or else open strings 
which intersect the boundaries of the cubic lattice. The shape of the strings 
will be Brownian with step size corresponding to the cell size of the lattice.


We have seen how relativistic MHD under the approximations discussed above can
be described in terms of magnetic flux tubes satisfying the Nambu-Goto equations of 
motion. In order to represent the continuous distribution of magnetic flux by 
a network of string, we gather together a flux $\Phi$ into ideal Nambu-Goto strings 
at positions $X^i(\bar\eta,\sigma)$, with \cite{olesen1}
\begin{equation}
B^{i}(\tau,\mathbf{x})=
\Phi\int d\sigma\frac{\partial X^i(\bar{\eta},\sigma)}{\partial\sigma}
\delta^{3}\left(\mathbf{x}-\mathbf{X}(\bar{\eta},\sigma)\right) 
\end{equation}

A few words need to be said about reconnection. In real fluids, magnetic flux
tubes interact and reconnections take place when the field lines cross each 
other. In ideal MHD there are no dissipative or viscous effects. Physical 
reconnection between field lines cannot take place without resistive effects and
therefore the topology of the magnetic field lines is frozen in the fluid and 
does not change with time.

Reconnection, which is a local process, is difficult to describe and the details
are not well understood. The efficiency of reconnection is not known and neither is how 
this process depend on the local geometry involved, like the relative 
inclination of the flux tubes. To answer this question, one has to go to 
numerical solutions of the underlying theory, that is MHD. 

In order to allow for reconnection to take place in our simulations we will take
a more simple approach to reconnection. In our model strings are allowed to
reconnect only if they pass through the same lattice point. When two strings 
meet, they intercommute with a probability $P$ and in this work we put $P=1$.
Reconnection here take place instantaneously between the discrete evolution 
time steps. This is physically reasonable since the reconnection timescale is
small compared with the evolution of the system. 

We use a simple estimate for the characteristic
length scale $\xi$ which is defined by 
\ben
\xi^2=V/L,
\een 
where $V$ is the volume of the 
lattice and $L$ is the total length of string in the box, excluding small 
loops.  In the Smith-Vilenkin algorithm it is in fact possible to have a loop 
of zero spatial extent, occupying just one lattice point, which travels at the
wave velocity.  If such a loop is formed it does not contribute to the magnetic 
field, and we do not count it in the calculation of the total length.  It is a
well-known feature of Nambu-Goto simulations that nearly all the string ends 
up in this kind of loop \cite{vil-shell,hind-kib}. In the MHD context 
we should probably not regard these loops as representing magnetic field 
lines at all: the energy is probably being dissipated in a very small-scale 
reconnection process. In any case, the fact that loops are generically produced 
so small underlines the fact that strings are very efficient at transferring 
power from large to small scales.


We have studied simulations of the evolution of the magnetic field on lattices 
with sizes $(128\delta)^{3}$ and $(256\delta)^{3}$, using a code originally 
developed by Sakellariadou \cite{SakVil88}, which implements both the 
Vachaspati-Vilenkin algorithm for the initial conditions and the Smith-Vilenkin 
alorithm for the evolution. Periodic boundary conditions
were used in all simulations and the evolution time were resticted to less than 
half the box size, since after this time causal influences have had time to 
propagate around the box.

Having a representation for the magnetic field in real space we use a three
dimensional Fast Fourier Transform algorithm to get a Fourier mode 
representation. The power spectrum for the magnetic field 
$|\mathbf{B}_{k}|^{2}$ can then be calculated at every time step by averaging
over the amplitudes of all Fourier modes with a wave number between $k$ and
$k+2\pi/\delta$. 

Previously it has been shown that networks of cosmic strings, modelled as Nambu-
Goto strings in Minkowski spacetime tend towords a scaling regime 
\cite{graham1}. This means that the characteristic length scale of the network
grows as $\sim c\,t$ where $t$ is Minkowski time.
The characteristic length scale for a primordial magnetic field $\xi$ does not 
grow with the horizon. Instead we expect the magnetic field to grow as 
$\sim c\,\bar{\eta}$. 

It is interesting to see if the magnetic field power spectra show scaling 
behaviour. In order to investigate this we express the power spectrum in terms 
of a scaling function $P$ which is defined through the following expression
\begin{equation}
|B(k,\bar{\eta})|^{2}=V \frac{\Phi^2 P(k\xi)}{\xi}.     
\label{e:scaling}
\end{equation}
Here $V$ is the volume of the box and its appearance is just a normalisation 
convention. This form for the scaling function ensures that 
that the fluctuations obey the scaling law
\begin{equation}
B^2 = \int \frac{d^{3}k}{(2\pi)^3} |B_{k}|^{2}\propto\xi^{-4}\Phi^2
\end{equation}
which is consistent with our picture of a coherent flux $\Phi$ in a region 
of size $\xi$.

Figure 1 shows the chacteristic length scale of the magnetic field versus 
$c\,\bar{\eta}$ for a typical ensemble.
The measured scaling function $P(k\xi)$ is displayed in Figure 2. The data was 
taken from 12 runs on a $(256\delta)^{3}$ lattice, for values of 
$c\,\bar{\eta}$ between $40-60$. It is seen that the power spectrum of the magnetic
field does reach a scaling regime. This means that the evolution of the network
will be self-simular with respect to $c\,\bar{\eta}$.

It is clear that the network does exhibit the property of scaling, with 
$\xi= x\!_*\cdot c\,\bar{\eta}$. The scaling amplitude $x\!_{*}$ can be obtained by 
looking at the ratio $\xi/\bar{\eta}$ towards the end of the simulation and it 
is roughly $x\!_{*}\simeq 0.3$ (see \cite{graham1} for a more accurate determination). 
Thus we can write
\begin{equation}
\xi=x\!_{*}\cdot c\,\bar{\eta}=x\!_{*}\int v_{A}(\eta)d\eta.  
\end{equation}
Unfortunately, we do not yet know how $v_A$ depends on the time parameter 
$\eta$ or the conformal time $\tau$.  The Alfv\'en velocity $v_A$ depends 
on $B$, which in turn depends on $\Phi$ and $\xi$ through (\ref{e:scaling}). 
All we can infer from the information at hand is a consistency relation:
if $\xi \propto \eta^r$ and $\Phi \propto \eta^s$, then $3r = s+1$.

The extra information we need comes from the covariance of ideal MHD 
under the scale transformation \cite{olesen2}
\ben
\bx \to l\bx, \quad
\tau \to l^{1-h} \tau, \quad
\bv \to l^h \bv, \quad
\bB \to l^h \bB,
\een
where $h$ is arbitrary.  One can show that under this transformation,
\ben
V^{-1} |\mathbf{B}_{k}|^{2} \to  l^{3+h} V^{-1} |\mathbf{B}_{k}|^{2}.
\een
If we define a function $\La(k,\tau)$ by
\ben
V^{-1} |\mathbf{B}_{k}|^{2} = \tau^{(3+h)/(1-h)} \La(k,\tau), 
\een
then we see that under the same transformation
\ben
\La(k,\tau) \to \La(k/l,\tau l^{1-h}) =\La(k,\tau), 
\een
from which we immediately infer that
\ben
\La(k,\tau) = \La(k\tau^{1/(1-h)}).
\een
Furthermore, if $|\mathbf{B}_{k}|^{2}$ behaves as $k^n$ 
as $k\to 0$, we have 
\ben 
|\mathbf{B}_{k}|^{2} \propto \tau^{(3+h+n)/(1-h)} k^n 
\een 
in that limit.  It is often assumed that the large-scale 
power is not affected by small-scale processes \cite{olesen2,son}, 
in which case $h=-n-3$, and we find the scaling laws (derived by 
the same authors)
\ben
\xi \propto \tau^{2/(n+5)}, \qquad \Phi \propto \tau^{(1-n)/(n+5)}.
\een
In the early universe individual particles move with relativistic velocities.
However, we expect that the bulk velocity of the fluid $v$ to be 
non-relativistic. Hence 
\ben
U^0 = qc \frac{\pa \tau}{\pa \eta} \simeq c,
\een
and given (\ref{e:qchoice}) we see that, on average, 
\ben
\eta = \sqrt{\langle{\dot\bx^2}\rangle} \tau /v_A,
\een
where $\langle{\dot\bx^2}\rangle$ is the mean square string velocity.  Simulations 
give this to be $0.36$ \cite{graham1}.

\section{Conclusions}

We have seen how the relativistic MHD equations, with a few 
reasonable assumptions, may  be recast as string-like equations for the 
motion of the flux lines.  This allows us enormous gains in dynamic 
range in the simulation of a random magnetic field in the early Universe,  
without being forced to the ideal limit, for we incorporate 
diffusivity by allowing reconnections between the strings.

The result is that we can understand the evolution of magnetic fields in
terms of the evolution of a network of strings, and we find that 
the power spectrum quickly evolves to a self-similar or scale-invariant 
form, with scale length $\xi$ increasing in time.  What this power law 
is we are unable to say:  ideal MHD predicts $\xi \propto \tau^{2/(n+5)}$, 
where $n$ is the low $k$ exponent of the power spectrum, but as we have 
departed from ideality by allowing reconnection, we cannot make a prediction.

The increase in scale length comes about by the strings straightening at the 
Alfv\'en velocity, while forming very small loops which can dissipate 
energy quickly.  This is the new feature that the string 
formulation brings to light:  strings transfer energy from large to small 
scales in an extremely efficient manner.  Thus, although the scale length 
increases, it is because power is preferentially lost from small scales. 
Whether it is fair to call this an inverse cascade is a matter of terminology.
What is clear is that the dynamics predicted by the string model of MHD 
is certainly not of the right kind to produce seeds for the galactic 
dynamo from magnetic fields created in the very early Universe.

There are of course many places where this line of argument is vulnerable. 
The model makes approximations which we have tried to highlight. Furthermore, 
our string simulations use special string configurations to make gains in
computational efficiency: the strings lie on a cubic lattice to start with, 
and one may be suspicious that this may introduce some artifice into the 
dynamics.  However, the propensity of a string network to scale is firmly 
believed, so we are confident that the magnetic field power spectrum will 
also scale.  What is probably not well approximated is the actual form of the 
power spectrum, which betrays the particularly string-like feature of a 
$k^{-1}$ tail, due to the fact that all the flux is held to be concentrated in 
a narrow tube.  Furthermore, the string model may be deficient in its 
description of helicity,
which is known to be extremely important in the development of true inverse 
cascades \cite{pouquet,corn,FieCar98}.  The helicity is represented by the 
linking number of the strings, but we are not able to incorporate a local 
contribution induced, for example, by twisted tubes of flux.
It may well be that we are 
missing some very important dynamics here.
We clearly need to check our results against a non-ideal MHD code, to 
see if the predicted self-similar dynamics emerges, and also to find 
the correct power law for the scale length.  This project is currently in hand.

We are extremely grateful to 
Mairi Sakellariadou for the use of her Minkowski space string code.  
We have also benefited from conversations with Axel Brandenburg, 
Carlo Barenghi, Richard Rijnbeek and Vladimir Semenov.
MH is supported by PPARC grant no.\ GR/L56305.
MC is supported by Centrala Studien\"amden (CSN).

\newpage

\begin{center}
\begin{figure}[p]
\scalebox{0.7}{\includegraphics{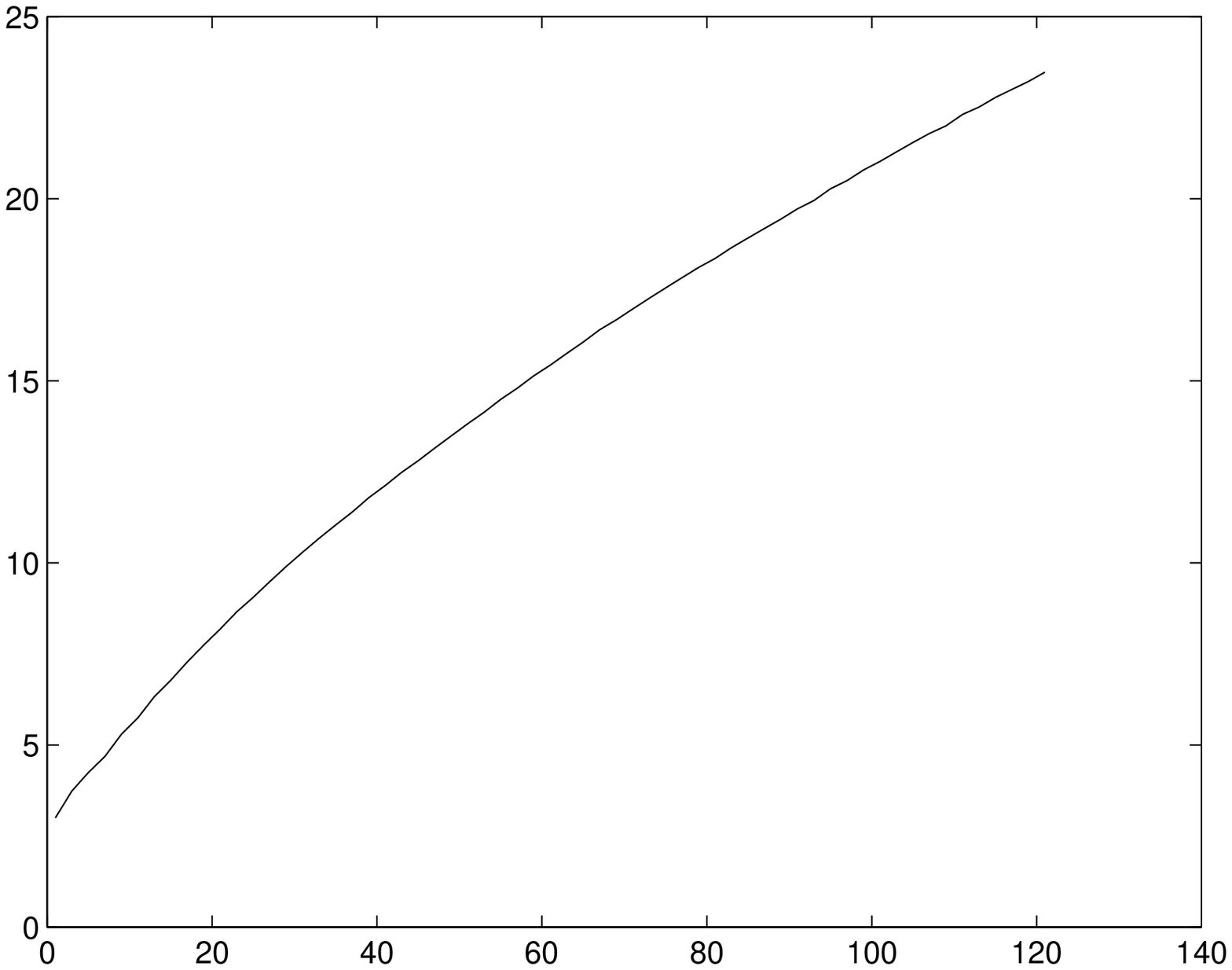}}
\caption{Correlation length $\xi$ versus $c\bar{\eta}$}
\end{figure}
\end{center}

\begin{center}
\begin{figure}[p]
\scalebox{0.7}{\includegraphics{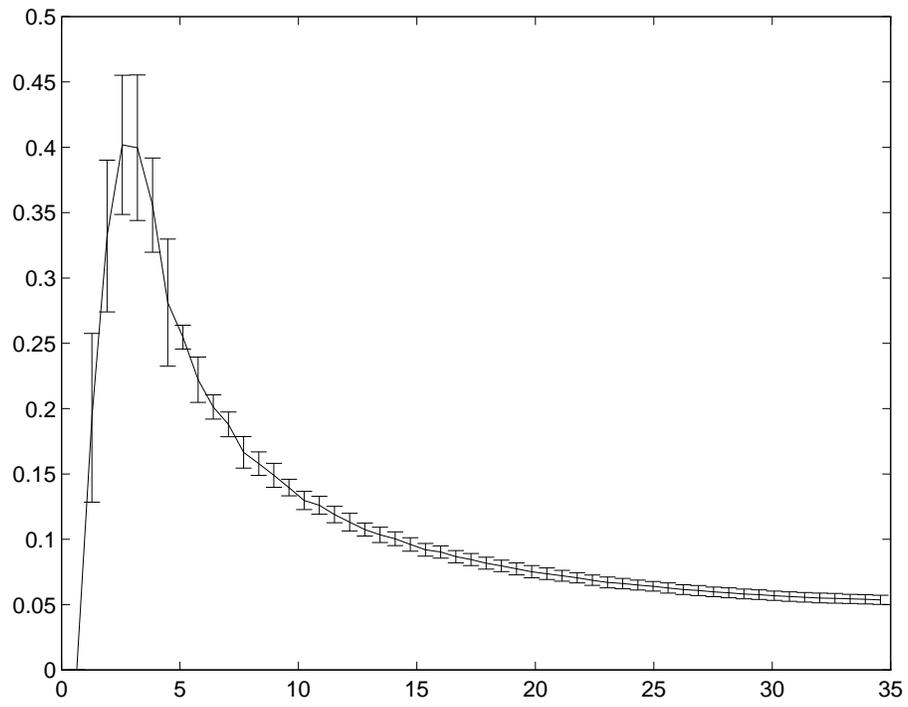}}
\caption{Scaling function $P(k\xi)$ versus $k\xi$}
\end{figure}
\end{center}

\end{document}